\documentclass[aps,prl,twocolumn,superscriptaddress]{revtex4-1}
\usepackage{amsmath,amsfonts,amssymb}
\usepackage{graphicx,float,calc}
\usepackage{color,bm}
\usepackage{ulem}
\usepackage{braket}
\usepackage[colorlinks,urlcolor=blue,citecolor=blue,linkcolor=blue]{hyperref}

\begin{document}
\title{Protecting Quantum Information via Many-Body Dynamical Localization}

\author{Ling-Zhi Tang}
\affiliation{Quantum Science Center of Guangdong-Hong Kong-Macao Greater Bay Area (Guangdong), Shenzhen 518045, China}
\affiliation{Key Laboratory of Atomic and Subatomic Structure and Quantum Control (Ministry of Education), Guangdong Basic Research Center of Excellence for Structure and Fundamental Interactions of Matter, South China Normal University, Guangzhou 510006, China}

\author{Dan-Wei Zhang}\email{danweizhang@m.scnu.edu.cn}
\affiliation{Key Laboratory of Atomic and Subatomic Structure and Quantum Control (Ministry of Education), Guangdong Basic Research Center of Excellence for Structure and Fundamental Interactions of Matter, South China Normal University, Guangzhou 510006, China}
\affiliation{Guangdong-Hong Kong Joint Laboratory of Quantum Matter, Frontier Research Institute for Physics, School of Physics, South China Normal University, Guangzhou 510006, China}

\author{Hai-Feng Yu}
\affiliation{Beijing Academy of Quantum Information Sciences, Beijing 100193, China}

\author{Z. D. Wang}
\email{zwang@hku.hk}
\affiliation{Department of Physics, and HK Institute of Quantum Science \& Technology, The University of Hong Kong, Pokfulam Road, Hong Kong, China}
\affiliation{Quantum Science Center of Guangdong-Hong Kong-Macao Greater Bay Area (Guangdong), Shenzhen 518045, China}

\begin{abstract}
Dynamically localized states in quantum many-body systems are fundamentally important in understanding quantum thermalization and have applications in quantum information processing. Here we explore many-body dynamical localization (MBDL) without disorders in a non-integrable quantum XY spin chain under periodical and quadratic kicks. We obtain the localization phase regimes with the MBDL and delocalized states and show dynamical observables to extract the phase regimes. For proper kick strengths in the MBDL phase, we reveal a local dynamical decoupling effect for persistent Rabi oscillation of certain spins. Furthermore, we propose the MBDL-protected quantum information at high temperatures, and present an analysis of the dynamical decoupling to obtain the required system parameters for quantum storage. Compared to other non-thermalized states, the disorder-free MBDL states require much fewer repetitions and resources, providing a promising way to protect and store quantum information robust against thermal noises.
\end{abstract}

\date{\today}
\maketitle

{\color{blue}\textit{Introduction.---}} Coherent and controllable many-body quantum dynamics plays an important role in quantum matter engineering and quantum information processing. As a consequence of thermalization, however, coherence in generic many-body quantum systems is fragile once they are excited to a finite energy density above ground states \cite{Rigol2008,DAlessio2016}. There are several ways to avoid thermalization, apart from integrability. 
One is the many-body localized (MBL) states induced by sufficiently strong disorders \cite{Anderson1958,Nandkishore2015,Abanin2019,Sierant2025}, and the other is quantum many-body scars \cite{Bernien2017,Turner2018,Bluvstein2021,Serbyn2021,PZhang2022}, which correspond to the strong and weak breaking of ergodicity, respectively. In MBL systems, all excitations are localized and thermalization is strongly suppressed, which can protect coherence of any initial states. The scar states also have a long coherence time as their dynamics are effectively restricted in a subspace. These localized states may provide resources for quantum information processing \cite{Huse2013,Serbyn2013a,Serbyn2014,Bahri2015,Yao2015,DWZhang2018,HDong2023,Naik2019,XMi2022,Schmid2024,Yarloo2024}.

Floquet engineering provides another versatile toolbox to understand thermalization and non-equilibrium phases in periodically driven quantum systems \cite{Alessio2014,Haldar2017,Moessner2017,Geier2021,Shimasaki2024,XMi2021,XZhang2022,HDong2024,Tiwari2024}. As a paradigm of quantum chaos, the quantum kicked rotor (QKR) can avoid chaotic dynamics due to the dynamical localization \cite{Fishman1982,CTian2004,CTian2011,Weidemann2021,Santhanam2022}, which is viewed as Anderson localization in momentum space and has been observed in ultracold atomic gases \cite{Moore1995,Lemarie2009,Lemarie2010,Manai2015}. Increasing efforts are made to study dynamically localized states under many-body interactions, such as interacting QKRs at the few-body limit \cite{Adachi1988,Boness2006,Rozenbaum2017,PQin2017,Notarnicola2018,Chicireanu2021,Paul2024}, integrable case \cite{Keser2016}, and mean-field level \cite{Shepelyansky1993,Gligoric2011,Lellouch2020}, which were experimentally observed using Bose-Einstein condensates in periodically pulsed optical lattices \cite{SeeToh2022,Cao2022}. Recent theoretical works suggest that the many-body dynamical localization (MBDL) persists in periodically kicked Bose gases in strongly interacting regimes \cite{Rylands2020,Vuatelet2021,Vuatelet2023,Fava2020}. The MBDL has just been observed in an interaction-tuable Bose gas \cite{YGuo2023}. Similar to the dynamical localization, the dynamical freezing of the magnetization for generic initial states in periodically driven spin chains due to emergent conservation laws is predicted \cite{Das2010,Haldar2018,Haldar2021}. Quantum spin chains with tunable disorders and driving fields have also been proposed to realize MBL and scar (confined) states \cite{Lazarides2015,Ponte2015,Bordia2017,Sierant2023,Kormos2016,SLiu2023,Verdel2023}. The periodically kicked Ising chains with exactly solvable dynamics can serve as a minimal model to study many-body quantum chaos \cite{Prosen2002,Bertini2018,Bertini2019,Khasseh2023,Sharma2024}.  Nevertheless, the MBDL in clean driven spin systems, which are amenable in quantum processors, and its related coherence protection remain unexplored.

In this work, we investigate the disorder-free MBDL in a non-integrable kicked XY spin chain and its application in protecting quantum information. For the quantum XY chain with a longitudinal field under periodic transverse-field kicks, we obtain the phase diagram with the MBDL and delocalized phases and show dynamical observables to extract the localization phase regimes. We uncover a local dynamical decoupling (DD) effect under proper kick strengths in the MBDL regime, which enables persistent Rabi oscillation of certain spins. Moreover, we demonstrate the MBDL-protected quantum information at high temperatures, and present an analysis of the local DD to obtain the system parameters for robust quantum storage. Our proposal is realizable with superconducting and Rydberg atom quantum simulators.

{\color{blue}\textit{Model.---}} We consider a kicked quantum spin chain of length $L$ described by the Hamiltonian:
\begin{equation}\label{Ht}
	H(t)=H_{\text{XY}}+KT \sum_{j=1}^{L}\left(j-j_0\right)^2 \sigma_j^z \sum_{n=1}^{N_T} \delta(t-n T),
\end{equation}
where $H_{\text{XY}}=\sum_{j=1}^{L} J\left(\sigma_j^{+} \sigma_{j+1}^{-}+\sigma_{j+1}^{+} \sigma_j^{-}\right)+\Omega \sigma_j^x$ is the static Hamiltonian with XY interaction between nearest-neighbour spins and a longitudinal field, $K$ and $T$ denote the strength and period of the transverse-field kicks with a quadratic potential centered at $j_0$, respectively. Here $\sigma_j^{\pm}=\sigma_j^{x}\pm i\sigma_j^{y}$ with $\sigma_j^{x,y,z}$ being the Pauli matrices of $j$-th spin, $J$ is the interaction strength, and $\Omega$ is the strength of the longitudinal field, which leads to coherent spin flips and breaks $U(1)$ symmetry. Notably, the quadratic form of the periodic kick introduces dynamical constraints on the spin chain, which tend to dynamically localize spins and compete with the XY interaction term. Other kinds of kick potentials, such as a linear potential, can also be chosen to induce the dynamical localization. We will emphasize in protection of quantum coherence raising from $\Omega$ via the MBDL in this kicked XY chain, which was not considered in kicked Ising chains \cite{Prosen2002,Bertini2018,Bertini2019,Khasseh2023,Sharma2024}. Hereafter we consider the high-frequency driving by setting $T=1/16$ and $\Omega=\hbar=1$ as the energy (frequency) unit.

For the kicked spin chain, the Floquet operator $U_F$ is
\begin{equation}\label{UF}
U_F=e^{-iKT \sum_{j=1}^{L}\left(j-j_0\right)^2 \sigma_j^z}e^{-iH_{\text{XY}}T}\equiv e^{-iH_FT},
\end{equation}
where $H_F=\frac{i}{T}\ln U_F$ is the effective Floquet Hamiltonian. The wave function at time $t=nT$ after $n$ kick cycles is $\ket{\psi(t)}=U_F^n \ket{\psi(0)}$.
As the model is non-integrable and $U(1)$-symmetry broken, we study the static and dynamical properties using the exact diagonalization \cite{Weinberg2019} for $L\leq14$ and use the time-dependent variational principle (TDVP) \cite{Fishman2022,SM} to study the dynamics of large systems up to $L=61$. Without loss of generality, we consider centrosymmetric spin chains by setting $j_0=(L+1)/2$ under open boundary condition.

\begin{figure}[tb]
	\centering
	\includegraphics[width=0.48\textwidth]{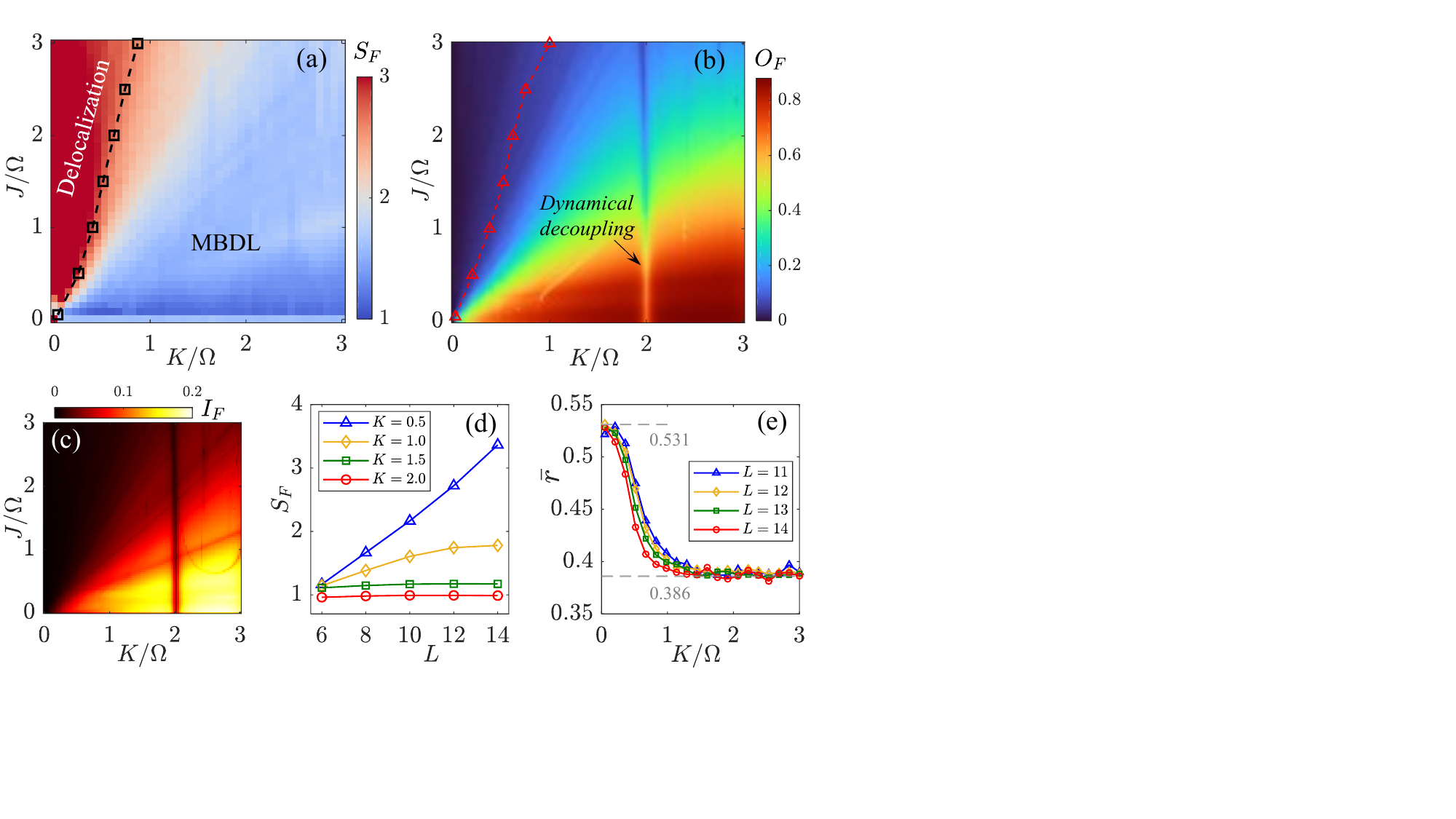}
	\caption{(Color online) Localization phase regimes on the $K$-$J$ parameter space obtained by (a) the averaged half-chain entanglement entropy $S_F$ of $L=12$, (b) the staggered magnetization $O_F$ and (c) the IPR $I_F$ averaging over all Floquet eigenstates of $L=11$. 
The dashed lines with squares and triangles for critical points in (a) and (b) represent the phase boundaries estimated by the finite-size scaling of $S_F$ and $O_F$ of Floquet eigenstates \cite{SM}, respectively. A proper kick gives rise to the DD of certain spins.
(d) $S_F$ versus $L$ for various $K$. (e) $\bar{r}$ as a function of $K$ for various $L$. Other parameter is $J=1$ in (c) and $J=2$ in (d,e).
	}\label{fig1}
\end{figure}

{\color{blue}\textit{Localization phase regimes.---}} We first study localization properties of many-body Floquet eigenstates $\{\ket{\psi^F_{\alpha}}\}$. They can be numerically obtained from the eigen equation $U_F \ket{\psi^{F}_\alpha}= e^{-i\epsilon_{\alpha}T} \ket{\psi^{F}_\alpha}$, where $\epsilon_{\alpha}\in[0,2\pi]$ are the folded quasi-energies with $\alpha=\{1,2,..,D\}$ and $D=2^L$ as the Hilbert-space dimension. To probe localization phases, we first compute the average half-chain entanglement entropy: $S_F=-\frac{1}{D} \sum_{\alpha=1}^D\operatorname{Tr}\left[\rho^F_{\alpha} \log \rho^F_{\alpha}\right]$, where $\rho^F_{\alpha}=\operatorname{Tr}_{j \leqslant L/2}[\ket{\psi^F_\alpha} \bra{\psi^F_\alpha}]$ with even $L$. The result of $S_F$ in Fig. \ref{fig1}(a) shows the localization phase regime in the $K$-$J$ parameter space. For weak (no) kick and strong interaction, the kicked spin chain is in the delocalization phase with large $S_F$ obeying the volume law, which corresponds to thermalization and loss of quantum information of initial states. As $K$ is increased, it can inter the MBDL (dynamical localization) phase with small $S_F$ obeying the area law in the interacting (noninteracting) case. Another quantity to reveal localization is the infinite-time staggered magnetization evolving from an initial N\'{e}el state $\ket{\psi(0)}=|\uparrow\downarrow\uparrow\downarrow...\rangle$, which can be calculated via the Floquet ensemble $\{\ket{\psi_\alpha^F}\}$ as $O_{F}=\frac{1}{L} \sum_{j=1}^{L}\sum_{\alpha=1}^{D}  (-1)^j |c_\alpha|^2 \bra{\psi^F_\alpha} \sigma^z_j  \ket{\psi^F_\alpha}$ \cite{Rigol2008}, with the coefficients $c_\alpha = \braket{\psi^F_\alpha | \psi(0)}$. Figure \ref{fig1}(b) shows the numerical result of $O_{F}$ in the $K$-$J$ plane, with similar phase regimes shown in Fig. \ref{fig1}(a). We also compute the average inverse participation ratio (IPR) $I_F = \frac{1}{D} \sum_{\alpha=1}^D \sum_{p=1}^D|\braket{p|\psi_\alpha^F}|^4$, where $|p\rangle$ denote product states in the computational basis. The numerical result in Fig. \ref{fig1}(c) shows small and large $I_F$ in the delocalized and MBDL phases, respectively, which agrees with that in Fig. \ref{fig1}(b).

In the localization regime, a proper kick strength can induce the DD of certain spins with persistent Rabi oscillation, such as two edge spins for $L=11$ and $K/\Omega=16\pi/25\thickapprox2.01$ in Figs. \ref{fig1}(b) and \ref{fig1}(c). Here the local DD leads to the decrease of $O_F$ and $I_F$ with respect to $K$, and will be discussed later. Fig. \ref{fig1}(d) further shows the area-law entanglement in the deep MBDL phase since only the area of the bipartitioning contributes to the entanglement \cite{Page1993,Luitz2015,Kjall2014,Pal2010}, which is different from the volume law in the delocalized phase. Figure \ref{fig1}(e) indicates the crossover from delocalization to the MBDL by the mean gap-ratio parameter $\bar{r}$ \cite{Alessio2014,GQZhang2020}. The level-spacing statistics of Floquet eigenstates is characterized
by $r_{\alpha}=\min\{\delta_{\alpha},\delta_{\alpha+1}\}/\max\{\delta_{\alpha},\delta_{\alpha+1}\}$, where $\delta_{\alpha}=\epsilon_{\alpha+1}-\epsilon_{\alpha}$ is the quasi-energy gap. We obtain $\bar{r}$ up to $L=14$ by averaging $r_{\alpha}$ over all Floquet eigenstates and over 20 realizations of spin chains with a center offset $j_0\rightarrow j_0+j_{\text{offset}}$ (to lift the level-spacing degeneracy), with $j_{\text{offset}}$ uniformly distributed in the small region $[-0.02,0.02]$. The MBDL phase is indicated by a Poissonian level-spacing distribution with $\bar{r}\approx0.386$, whereas presents a Wigner-Dyson distribution with $\bar{r}\approx0.53$ in the delocalized phase.

Similar to the MBL transition \cite{Nandkishore2015,Abanin2019,Sierant2025}, the transition between the MBDL and delocalized (thermal) phases is a dynamical phase transition. The understanding of the transition is still notoriously incomplete due to the lack of analytical treatments and the inherent limitations in numerical simulations. Here we focus on the phase regimes for the MBDL and estimate the phase boundaries by the finite-size scaling of $S_F$ and $O_{F}$ for small system sizes with $L\leq14$ by the exact diagonalization (see the Supplementary Material (SM) \cite{SM}). The numerical results give roughly consistent phase boundaries in Figs. \ref{fig1}(a) and \ref{fig1}(b). It remains an important task to capture the criticality near the MBDL-delocalization transition \cite{Rispoli2019}. However, our results pinpoint large parameter regions for the MBDL phase in the kicked spin chain.


%
{\color{blue}\textit{Dynamical observables.---}} The (de)localization can be observed via the stroboscopic dynamics of the kicked spin chain. For an initial product state in experiments, a direct observation is the spin-up (excitation) density distribution $P_{\uparrow}(j,t)=\left\langle\psi(t)\left|\frac{\sigma^z_j+\mathbb{I}}{2}\right| \psi(t)\right\rangle$ at the stroboscopic time $t=nT$, where $\mathbb{I}$ is the $2\times2$ identity matrix. The results of $P_{\uparrow}(j,t)$ for different initial states show similar dynamics \cite{SM}: The spin chain tends to thermalize in the static case. Turning on periodical kicks up to the MBDL regime, the quantum coherence of initial states preserves after the long-time evolution. This disorder-free non-thermalized dynamical phase is due to the kinetic constriction by the quadratic kick potential. The MBDL can be broken down and the thermalization dynamics dominates as $J$ is increased.

\begin{figure}[tb]
	\centering
	\includegraphics[width=0.47\textwidth]{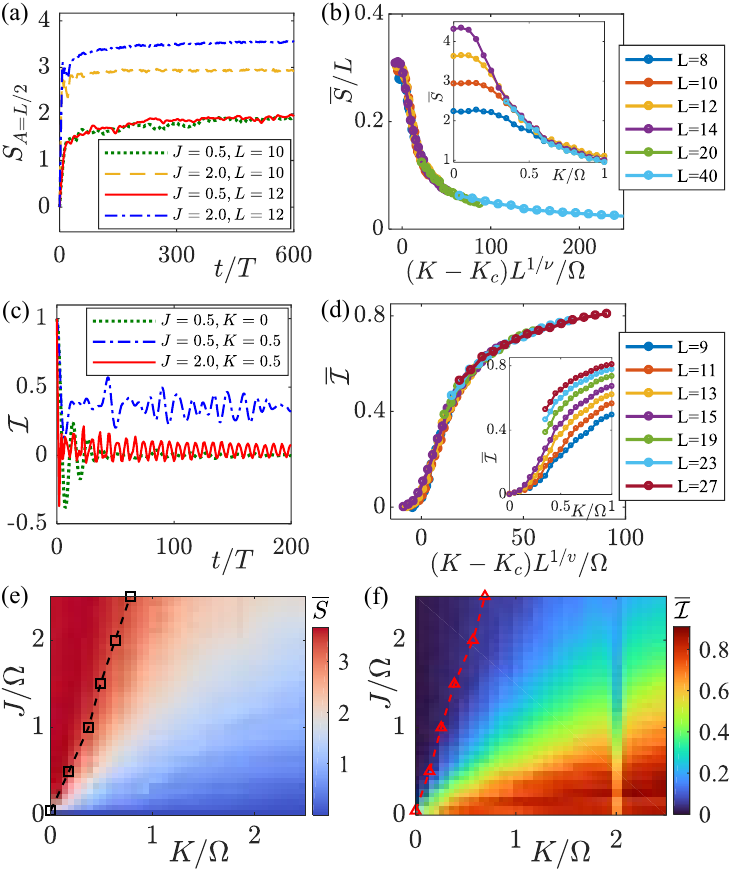}
	\caption{(Color online) (a) Time evolution of subsystem entanglement $S_{A=L/2}$ for typical parameters. (b) Finite-size scaling of $\bar{S}$. (c) Time evolution of $\mathcal{I}$ for typical parameters. (d) Finite-size scaling of $\overline{\mathcal{I}}$. Insets in (b) and (d) show $\bar{S}$ and $\overline{\mathcal{I}}$ versus $K$ for various $L$, respectively. (e,f) Localization phase regimes with rough boundaries extracted by $\bar{S}$ and $\overline{\mathcal{I}}$, respectively. Other parameters are $L=12$ in (e), $L=11$ in (f) and $J/\Omega=0.5$ in (b,d), and the initial N\'eel state is $\ket{\psi(0)}=|\uparrow\downarrow\uparrow\downarrow...\rangle$ in (a-h). The numerical results for $L>14$ are obtained by the TDVP in the localized regime where the entanglement is low \cite{Fishman2022,SM}.
	}\label{fig2}
\end{figure}

To further indicate the MBDL and extract the localization phase regimes in Figs. \ref{fig1}(a) and \ref{fig1}(b), we propose to measure the dynamical observables corresponding to $S_F$ and $O_F$. Firstly, one can prepare an initial N\'{e}el state and measure the evolution of subsystem entanglement entropy $S_A(t)=-\operatorname{Tr}\left[\rho_{A}(t) \log \rho_{A}(t)\right]$ \cite{Kaufman2016,Rispoli2019,HDong2023,Choi2023}, where $\rho_{A}(t)=\operatorname{Tr}_{j \leq A}[\ket{\psi(t)} \bra{\psi(t)}]$ is the reduced density matrix of subsystem-A. By performing tomography, one can measure $\rho_{A}$ after $n$ kicks to obtain $S_A(t)$. Figure \ref{fig2}(a) shows distinct saturation behaviors of the half-chain entanglement $S_{A=L/2}$ for various $L$ in the MBDL and delocalized phases. In finite localized (delocalized) systems, $S_{A=L/2}(t)$ approaches to saturate at late times to values independent (linearly dependent) of $L$ \cite{Fagotti2008,Bardarson2012,Serbyn2013b,Hart2021,Lim2024}. Moreover, we can use the long-time-averaged entanglement entropy $\overline{S}$ (averaging over the time interval from $t=400T$ to $600T$ in simulations) to indicate the MBDL, similar as that in Ref. \cite{Verdel2023}. The numerical results of $\overline{S}$ versus $K$ for $J=0.5$ and various $L$ are shown in Fig. \ref{fig2}(b). A finite-size scaling analysis of $\overline{S}$ yields the estimated critical point at $K_c=0.18$ and critical exponent $\nu=0.65$ for the considered parameters.

Another observable is the generalized imbalance $\mathcal{I}(t)=\frac{1}{L}\sum_{j}(-1)^{j+1}P_{\uparrow}(j,t)$, which approaches to the value of $O_F$ in the infinite-time limit. The results of $\mathcal{I}(t)$ for finite chains are shown in Fig. \ref{fig2}(c). In the MBDL (delocalized) phase when $L\rightarrow\infty$, $\mathcal{I}$ remains finite (becomes vanishing) after long time evolution. The long-time-averaged imbalance $\overline{\mathcal{I}}$ (averaging over the time interval from $t=100T$ to $200T$ in simulations) can also be used to detect the MBDL and estimate the critical point. As shown in Fig. \ref{fig2}(d), the finite-size scaling of $\overline{\mathcal{I}}$ yields $K_c=0.14$ and $\nu=0.70$, which are close to those in Fig. \ref{fig2}(b). Finally, we plot the results of $\overline{S}$ and $\overline{\mathcal{I}}$ in the $K$-$J$ plane in Figs. \ref{fig2}(e) and \ref{fig2}(f), respectively, for the same system sizes as those in Figs. \ref{fig1}(a) and \ref{fig1}(b). By the finite-size scaling of $\overline{S}$ and $\overline{\mathcal{I}}$ for various values of $J$, we obtain the phase boundaries. These results for different static and dynamical quantities give consistent localization phase regimes and rough boundaries.

\begin{figure}[tb]
	\centering
	\includegraphics[width=0.48\textwidth]{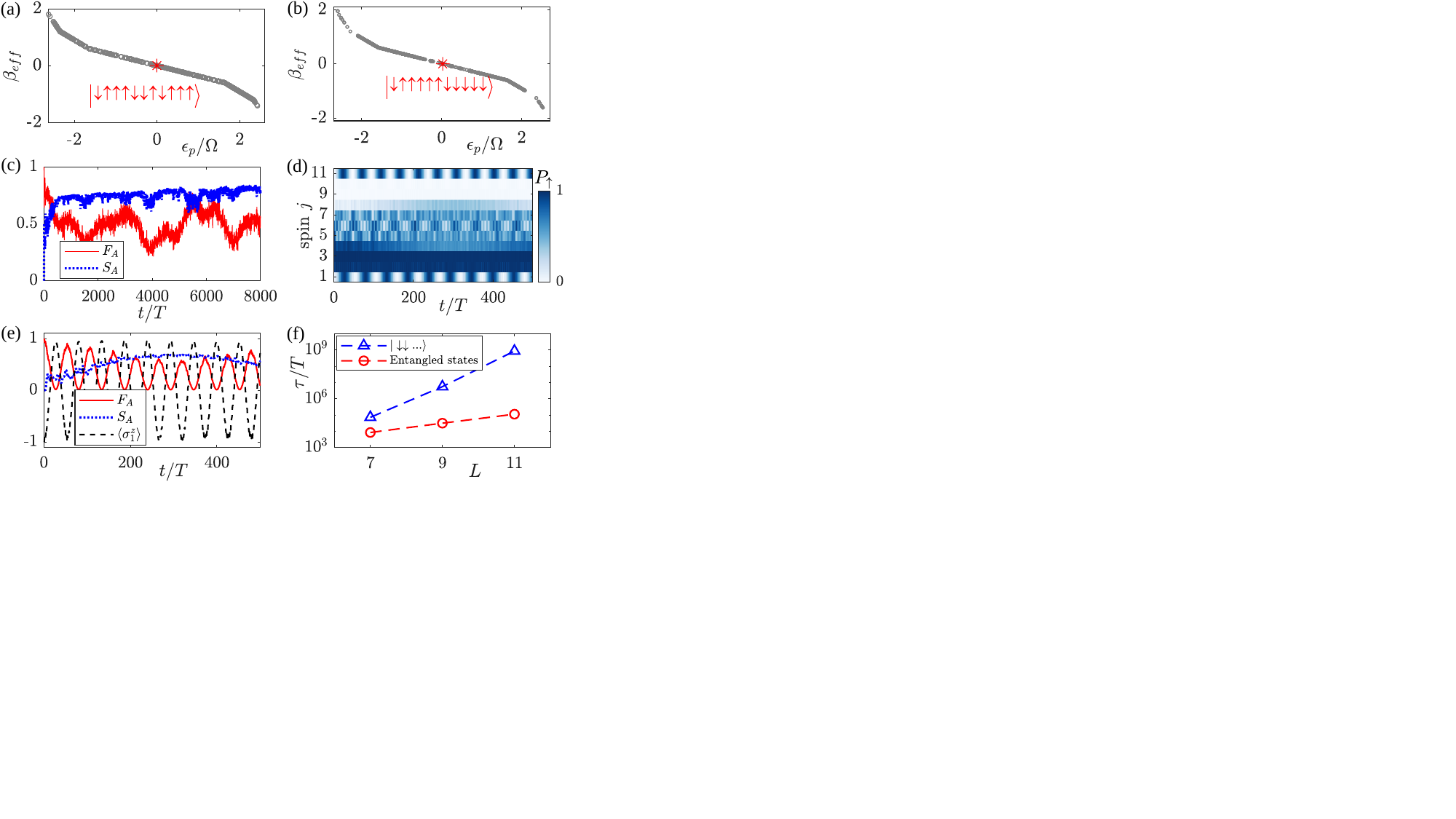}
	\caption{(Color online) $\beta_{\text{eff}}$ as a function of $\epsilon_p$ for (a) MBDL with $K=1$; and (b) DD with $K=16\pi/25$. The initial high-temperature states with $\beta_{\text{eff}}\approx0$ are denoted in red. (c) Time evolution of $F_A$ and $S_A$ (with $A=4$) for the initial state in (a). Time evolution of (d) $P_{\uparrow}$; and (e) $F_A$, $S_A$ and $\langle\sigma_1^{z}\rangle$ for the initial state in (b). (f) Lifetime $\tau$ of the $\langle\sigma_1^{z}\rangle$-oscillation versus $L$ for the initial vacuum and entangled states \cite{SM}. Other parameters are $J=0.5$ and $L=11$.
	}\label{fig3}
\end{figure}

{\color{blue}\textit{MBDL-protected quantum information.---}} In the deep MBDL phase, arbitrary states in the energy spectrum are non-thermalized. Due to this strong breaking of ergodicity, the quantum information can be robustly encoded with initial MBDL states. For a generic initial state $|\psi(0)\rangle$, its effective inverse temperature $\beta_{\text{eff}}$ can be obtained by solving $\operatorname{Tr}[H_F(\ket{\psi(0)}\bra{\psi(0)}-\rho_\beta)]=0$, where $\rho_{\beta}=e^{-\beta_{\text{eff}} H_F}/\operatorname{Tr}(e^{-\beta_{\text{eff}} H_F})$ is the thermal density matrix \cite{Banuls2011,FChen2021}. We plot $\beta_{\text{eff}}$ as a function of quasi-energies of all product states $\epsilon_p=\bra{\psi(0)}H_F\ket{\psi(0)}$ for the system of $L=11$ in the MBDL phase and local DD case in Fig. \ref{fig3}(a) and \ref{fig3}(b), respectively.
To show the protection of quantum information at high temperatures, we can consider the initial product states with $\beta_{\text{eff}}\approx0$ near the center of the energy spectra. The non-thermalization dynamics of initial MBDL states can be observed from the long-time coherent evolution of spin density distributions, with an example shown in Fig. \ref{fig3}(d) (and the SM \cite{SM}). We can study additional information on the coherent dynamics beyond local observables. We compute the time evolution of fidelity $F_A(t)=\left[\operatorname{Tr}\sqrt{\sqrt{\rho_A(t)}\rho_A(0)\sqrt{\rho_A(t)}}\right]^2$ and entanglement entropy $S_A(t)$ of subsystem-A with $A=4$ for the high-temperature initial states in Figs. \ref{fig3}(c) and \ref{fig3}(e). The subsystem fidelity with high values and low entanglement entropy persists in the long-time evolution. Note that some spins in the center of the chain could be non-thermalized and loss local information. To improve the fidelity of initial states, one can increase the ratio between the kick and interaction strengths $K/J$ (see Fig. S8 in the SM \cite{SM}).

For the local DD case in Fig. \ref{fig3}(d), one can find the persistent Rabi oscillation in $P_{\uparrow}(j,t)$ of $j=1,L$ for $L=11$. In this case, the two edge spins of arbitrary initial states are effectively decoupled and behave as free spins with nearly the same Rabi oscillation, as long as the system is in the MBDL phase. Similar coherent dynamics and the DD of edge spins for other initial states are shown in the SM \cite{SM}. Note that the DD effect will be destroyed if $J$ is increased up to the delocalization regime. Interestingly, the revivals in the local density dynamics in Fig. \ref{fig3}(d), as well as $\langle\sigma^z_1(t)\rangle$ and $F_A(t)$ in Fig. \ref{fig3}(e) imply that the quantum information of initial states in the DD case can be better restored. Moreover, we estimate the lifetime $\tau$ of the oscillation of $\langle\sigma^z_1(t)\rangle$ under the DD effect \cite{SM}. The results of $\tau$ versus $L$ for vacuum ($|\downarrow\downarrow...\rangle$) and entangled initial states shown in Fig. \ref{fig3}(f). Here the exponential increase of the lifetime with the system size indicates strong non-thermalization. Thus, the MBDL states with the DD effect provide a promising way of quantum storage robust against to thermal noises and even far away equilibrium. Notably, the long-time coherent dynamics is exhibited for generic initial MBDL states, such as entangled states shown in the SM \cite{SM}. Compared to the MBL and scar states, the MBDL states in this kicked spin chain do not require strong disorder average and special quantum state preparation.

\begin{figure}[tb]
	\centering
	\includegraphics[width=0.48\textwidth]{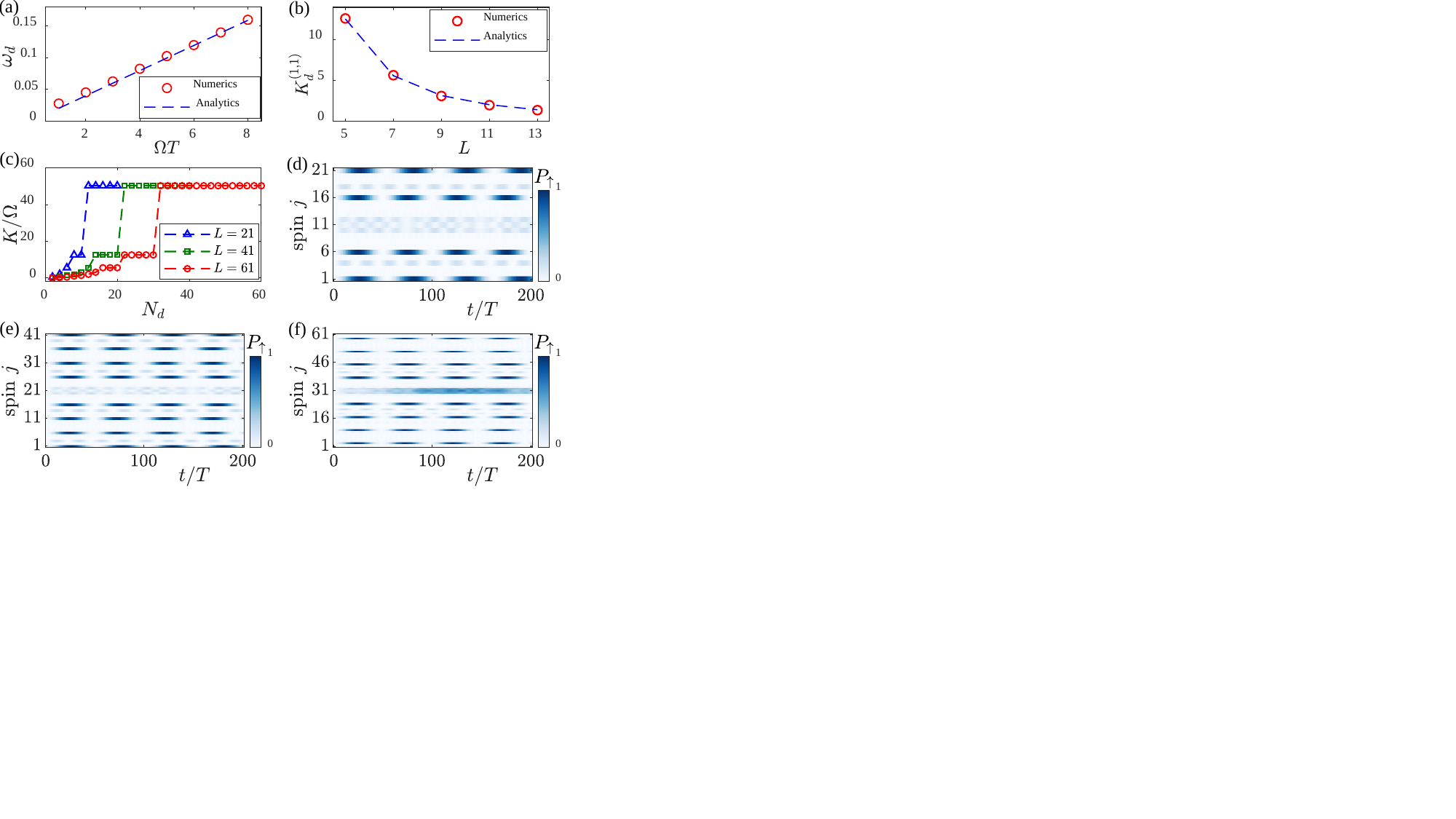}
	\caption{(Color online) (a) $\omega_d$ with respect to $\Omega$ for the DD of edge spins for $L=11$. (b) $K_d^{(1,1)}$ with respect to $L$. (c) $K$ versus $N_d$ for $L=21,41,61$.
Time evolution of $P_{\uparrow}(j,t)$ in the DD case of (d) $L=21,N_d=4,K=16\pi/25$, (e) $L=41,N_d=8,K=16\pi/25$, and (f) $L=61,N_d=8,K=16\pi/49$.
	}\label{fig4}
\end{figure}

{\color{blue}\textit{Analysis of local DD.---}} We proceed to present an analysis of the local DD to obtain the required system parameters. For the $j$-th decoupled spin in the MBDL regime, its effective coupling strength is vanishing, corresponding to the local Floquet operator  $U^{(j)}_F=e^{-i\tilde{K} \sigma_j^z} e^{-i \Omega T \sigma_j^x}$, where $\tilde{K}=KT(j-j_0)^2$. When $\tilde{K}=m\pi$ with $m$ being positive integers for certain kick strengths, the operator reduces to $U^{(j)}_F= (-1)^m e^{-i \Omega T \sigma_j^x}$. This implies the Rabi oscillation of the effectively decoupled spin with the frequency $\omega_d=\Omega T/\pi$. The analytical expression of the oscillation frequency is consistent of the numerical result, as shown in Fig. \ref{fig4}(a). The kick strength required for the DD of $j$-th spin is $K_d^{(j,m)}=m \pi/T(j-j_0)^2$. For the centrosymmetric chain of odd spins considered here, the local DD occurs with pairs of spins, such as $j=1,L$ for $L=11$ shown in Fig. \ref{fig3}(d) and Fig. \ref{fig4}. We numerically extract $K_d^{(1,1)}$ for various $L$ from the local dynamics and find well agreement with the analytical expression  $K_d^{(1,1)}=4\pi/T(L-1)^2$, as shown in Fig. \ref{fig4}(b). We further obtain the following relation between the kick strength and the DD spin number $N_d$ of system size $L$ \cite{SM}:
\begin{equation}
	K(N_d,L)=\frac{\pi}{T \lfloor(L-1)/N_d\rfloor ^2},
\end{equation}
where $\lfloor \cdots\rfloor$ is the floor function. The results for larger systems in Figs. \ref{fig4}(c) show that the number of DD spins can be increased up to $N_d=L-1$ by increasing the kick strength. We further verify this relation via the dynamics of the spin chains with $L=21,41,61$ by the TDVP, as shown in Figs. \ref{fig4}(d-f), respectively.

{\color{blue}\textit{Discussions and conclusions.---}} Before concluding, we discuss experimental realization of the kicked XY quantum spin chain. The first system is the superconducting quantum simulators, consisting of 1D arrays of controllable qubits with tunable nearest-neighbour XY coupling \cite{XLi2023,HDong2024,XTan2021}. One can individually and dynamically tune local potentials to realize the static and periodical kick terms along the $x$- and $z$-axis in Eq. (\ref{Ht}), similar as those in experiments in Ref. \cite{HDong2024}. Alternatively, one can realize the Floquet unitary $U_F$ given by Eq. (\ref{UF}) in a digital way using tunable quantum gates on superconducting qubits \cite{XMi2021,XZhang2022}. The second platform is the 1D driven Rydberg atom array \cite{Bluvstein2021,CChen2023}. One can encode the effective spin-1/2 in a pair of opposite-parity Rydberg states and realize the XY spin Hamiltonian via the resonant dipole-dipole interactions \cite{CChen2023}. The kicked potential may be produced by applying pulsed laser beams with spatially dependent light shifts \cite{Bluvstein2021,CChen2023}. Both superconducting and Rydberg quantum simulators are programmable with controllable many-body dynamics to detect the MBDL phase and the related coherence protection.

In summary, we have explored the disorder-free MBDL in the non-integrable quantum XY chain under periodical kicks. We have revealed the localization phase regimes with dynamical observables for the MBDL, and uncovered the local DD effect for persistent Rabi oscillation of certain spins. Furthermore, we have proposed the MBDL-protected quantum information at high temperatures, and presented an analysis of the DD to obtain the required system parameters for quantum storage. The MBDL states in spin systems offer a promising platform for studying quantum thermalization and quantum information processing robust against thermal noises.

\acknowledgments{The authors acknowledge stimulating discussions with Chushun Tian, Dong-Ling Deng, Zhen-Yu Wang, and Shi-Liang Zhu. This work was supported by the National Natural Science Foundation of China (Grants No. 12174126 and No. 92365206), Innovation Program for Quantum Science and Technology (No. 2021ZD0301802), the Guangdong Basic and Applied Basic Research Foundation (Grant No. 2024B1515020018), the Science and Technology Program of Guangzhou (Grant No. 2024A04J3004), and Quantum Science Center of Guangdong-Hong Kong-Macau Great Bay Area.}

\bibliography{reference}

\clearpage

\section{Supplemental Materials}

\subsection{A. The TDVP method}
We review some details of the time-dependent variational principle (TDVP) method \cite{Fishman2022}, which is useful for studying the dynamics of one-dimensional strongly correlated systems with large system sizes. We also test the accuracy of the TDVP method with different bond dimension and benchmark the results with those from the exact diagonalization (ED). In general, one can transform a many-body quantum state $\ket{\Psi}$ into a matrix product state (MPS) as \cite{Fishman2022}
\begin{equation}
	\begin{aligned}|\Psi\rangle=& \sum_{j_{1}, \ldots, j_{L}} \sum_{a_{0}, a_{1}, \ldots, a_{L}} \operatorname{Tr}\left(M_{a_{0}, a_{1}}^{j_{1}} M_{a_{1}, a_{2}}^{j_{2}} \cdots\right.\\ &\left.\times M_{a_{L-2}, a_{L-1}}^{j_{L-1}} M_{a_{L-1}, a_{L}}^{j_{L}}\right)\left|j_{1}, j_{2}, \ldots, j_{L-1}, j_{L}\right\rangle, \end{aligned}
\end{equation}
where $M_{a_{n-1},a_n}^{j_n}$ denotes tensors of rank-three with $a_n$ standing virtual index connecting neighboring sites, and $j_n$ marks local state index. The canonical form of the MPS can be obtained through the gauge freedom:
\begin{equation}
	|\Psi\rangle=\sum_{\alpha=1}^{\chi_{n}} \Lambda_{\alpha \alpha}^{[n]}\left|l_{\alpha}^{[n]}\right\rangle\left|r_{\alpha}^{[n]}\right\rangle,
\end{equation}
where $\Gamma^{[n]}$ denotes diagonal matrix with Schmidt values of subsection $\{l,r\}$, $n$ marks the canonical center, $\chi$ stands bond dimension which corresponds to summary of Schmidt values. The left (right) part of Schmidt states is $\left|l_{\alpha}^{[n]}\right\rangle=\sum_{j_{1}, \ldots, j_{n}}\left(A^{j_{1}} \cdots A^{j_{n}}\right)_{\alpha}\left|j_{1}, \ldots, j_{n}\right\rangle$ $\left( \left|r_{\alpha}^{[n]}\right\rangle=\sum_{j_{n+1}, \ldots, j_{L}}\left(B^{j_{n+1}} \cdots B^{j_{L}}\right)_{\beta}\left|j_{n+1}, \ldots, j_{L}\right\rangle\right)$ with $A^{j_i}$ ($B^{j_i}$) denoting the left (right) canonical form of $M^{j_i}$.

\begin{figure}[tb]
	\centering
	\includegraphics[width=0.45\textwidth]{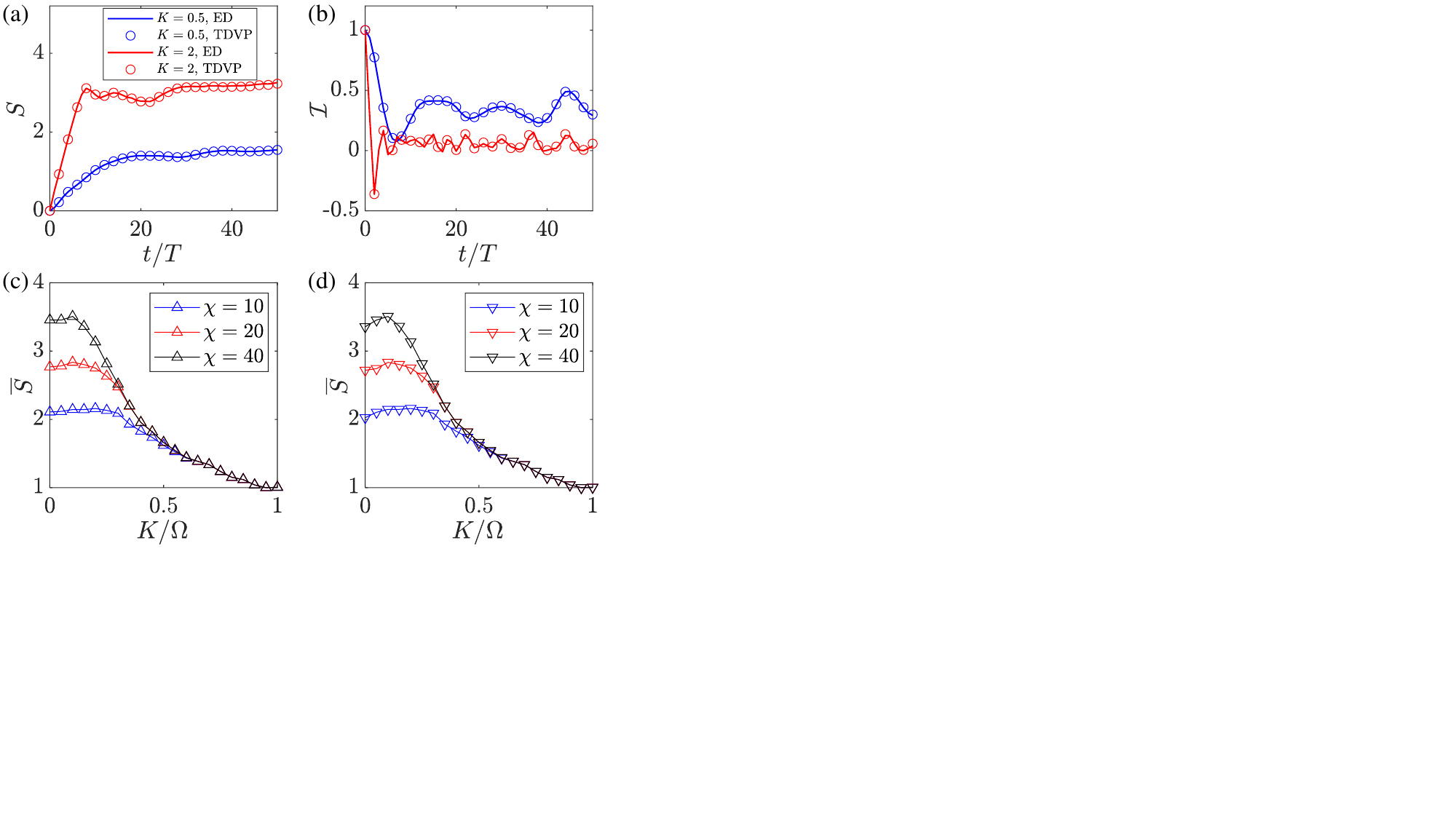}
	\caption{(Color online) Time evolution of (a) Entanglement entropy and (b) imbalance calculated by ED (solid lines) and TDVP (circles) for $K=0.5,2$. Saturation entanglement $\overline{S}$ for (c) $L=20$ and (d) $L=40$ as functions of $K$ for different bond dimensions $\chi=10,20,40$. Other parameters are $J/\Omega=0.5$ and initial N\'eel states are chosen. The systems sizes fixed $L=12$ in (a) and $L=11$ in (b).
	}\label{figS1}
\end{figure}

One can abstract TDVP method as follow:
(1) Obtain the initial MPS $\Psi_0$:
\begin{equation}
	\left|\Psi_{0}\right\rangle=\sum_{\alpha, j_{n}, j_{n+1}, \beta} M^{j_{n}, j_{n+1}}\left|l_{\alpha}^{[n-1]}\right\rangle\left|j_{n}\right\rangle\left|j_{n+1}\right\rangle\left|r_{\beta}^{[n+1]}\right\rangle.
\end{equation}
In each calculation, improve $M^{j_{n}, j_{n+1}}=A^{j_{n}} \Lambda^{[n]} B^{j_{n+1}}$ while fixing the tensors of $\ket{l^{[n-1]}}$ and $\ket{r^{[n+1]}}$.
(2) Calculate effective Hamiltonian $\hat{H}_{eff}$ in basis of $\ket{l_\alpha^{[n-1]} j_n j_{n+1} r_\beta^{[n+1]}}$.
(3) Calculate matrix exponential of the MPS: $\tilde{M}^{j_{n}, j_{n+1}}=\exp \left(-i \hat{H}_{\mathrm{eff}} \tau\right) M^{j_{n}, j_{n+1}}$ with evolved time $\tau$ \cite{Haegeman2016}.
(4) Left sweep: perform singular value decomposition (SVD) to $\tilde{M}^{j_n,j_{n+1}}=\tilde{A}^{j_n} \tilde{\Lambda}^{[n]} \tilde{B}^{j_{n+1}}$ to obtain tensor $A^{j_n}=\tilde{A}^{j_n}$; Backward evolution of $\tilde{\Lambda}^{[n]} \tilde{{B}}^{j_{n+1}}$ with evolved time $\tau$; Right sweep: perform SVD to $\tilde{M}^{j_n,j_{n+1}}=\tilde{A}^{j_n} \tilde{\Lambda}^{[n]} \tilde{B}^{j_{n+1}}$ to obtain tensor $B^{j_{n+1}}=\tilde{B}^{j_{n+1}}$; Backward evolution of $\tilde{A}^{j_{n}} \tilde{\Lambda}[n]$ with evolved time $\tau$.
(5) Back to second step and repeat the sweeping processes until target evolved time.

In Figs. \ref{figS1}(a) and \ref{figS1}(b), we numerically calculate the time evolution of the subsystem entanglement entropy $S$ and the imbalance $\mathcal{I}$, respectively. The numerical results for different $K$ and fixed $J=0.5$ are obtained through both TDVP and ED methods. The accuracy of the TDVP method is related to the bond dimension. Thus, we test the accuracy of the TDVP for larger system (e.g. $L=20,40$) after long evolved time (e.g. $t/T=600$). Figures \ref{figS1}(c) and \ref{figS1}(d) show $\bar{S}$ as functions of $K$ for different bond dimension $\chi$ and $L=20,40$, respectively. The results of region $K\in(0,0.35)$ diverse due to the large entanglement of evolving states, which indicate the system in the delocalized (thermal) phase. In the remained region of $K\in(0.35,1)$, $\bar{S}$ can saturate to a finite value by increasing $\chi$ as the entanglement is still low. In this case, the system is in the MBDL phase and we take the data.

\begin{figure*}[tb]
	\centering
	\includegraphics[width=0.85\textwidth]{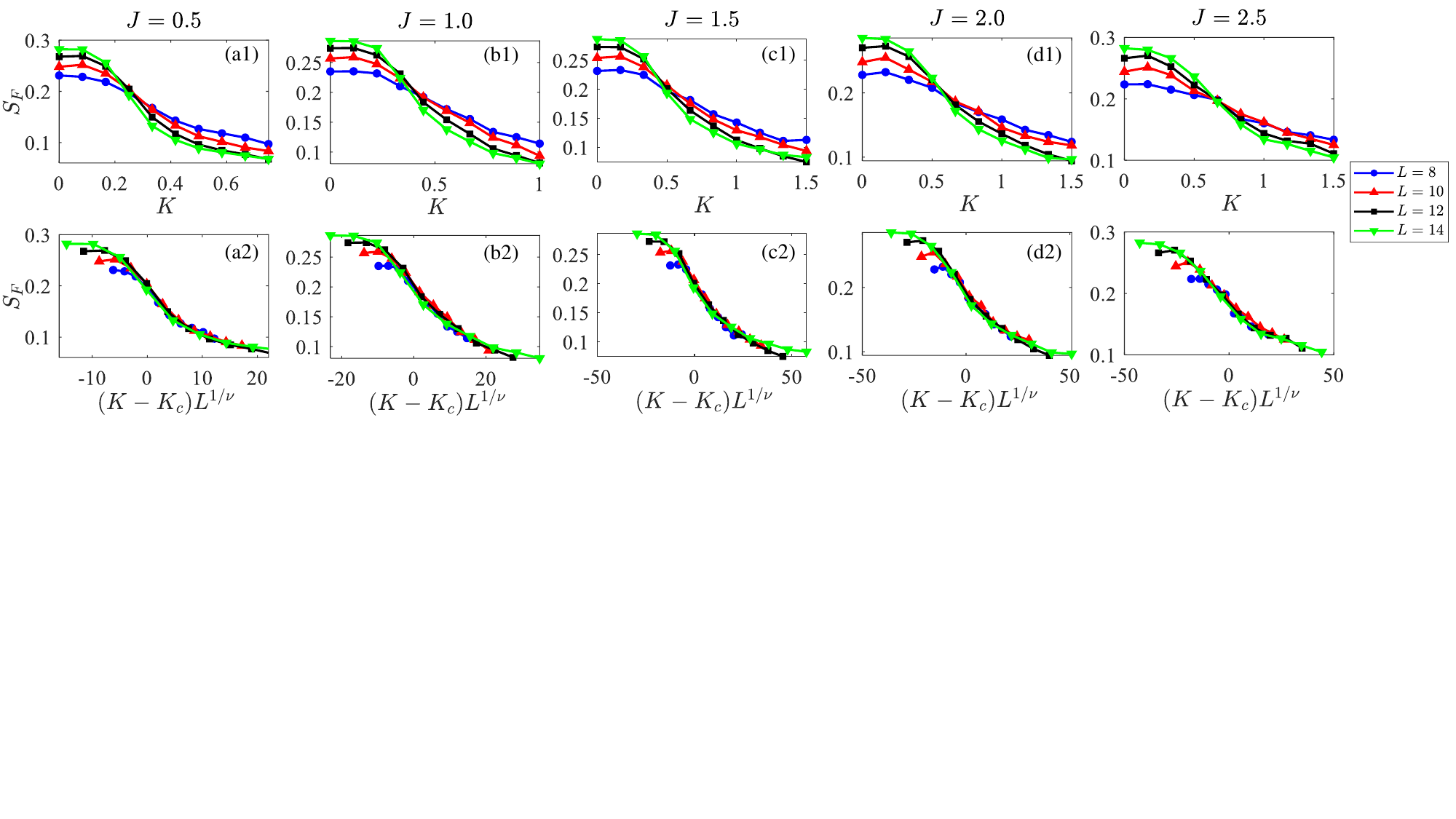}
	\caption{(Color online) Finite-size scaling of $S_F$ to estimate the critical points $K_c$ for $L=8,10,12,14$. The lower column shows $S_F$ after the data collapse for $J=\{0.5,1.0,1.5,2.0,2.5\}$ with $K_c\approx\{0.25,0.40,0.51,0.62,0.74\}$ and critical exponent $\nu\approx0.65$ in (a-e), respectively.
	}\label{figS2}
\end{figure*}

\begin{figure*}[tb]
	\centering
	\includegraphics[width=0.85\textwidth]{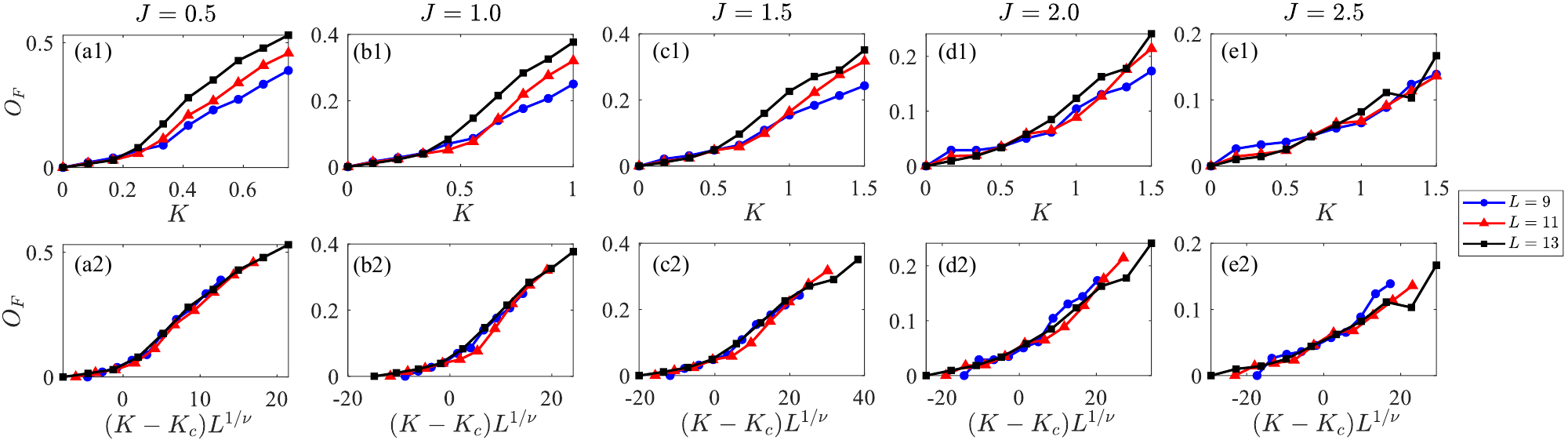}
	\caption{(Color online) Finite-size scaling of $O_F$ to estimate the critical points $K_c$ for $L=9,11,13$. The lower column shows $O_F$ after the data collapse for $J=\{0.5,1.0,1.5,2.0,2.5\}$ with $K_c\approx\{0.2,0.38,0.52,0.62,0.75\}$ and critical exponent $\nu\approx0.70$ in (a-e), respectively.
	}\label{figS3}
\end{figure*}

\subsection{B. Finite-size scaling to estimate phase boundaries}

We numerically estimate the phase boundary between the MBDL and delocalization phase through the finite-size scaling analysis of
infinite-time staggered magnetization $O_F$ and half-chain entanglement entropy $S_F$ defined in the main text~\cite{Page1993,Luitz2015,Kjall2014,Pal2010}.
For fully thermal Floquet states in our system, the entanglement entropy is given by the Page value $S_F^{\text{Page}}=(L\ln 2 -1)/2+O(L)$~\cite{Page1993}, which indicates that the value of entanglement entropy increases linearly with system size $L$, i.e., the volume law. On the other hand, these eigenstates in MBL systems are partly entangled, hence the entanglement entropy $S_F\sim O(L^0)$ obeys the area law. Figures \ref{figS2}(a1-e1) show the site-averaged half-chain entanglement entropy $S_F/L$ of Floquet eigenstates of $U_F$ as functions of the kick strength $K$ for different system sizes $L$ and $J$. For $J=0.5$ in Fig. \ref{figS2}(a1), $S_F/L$ obeys the volume law at weak kick strength, which indicates the thermalization in the system. As $K$ is increased, the curves of $S_F/L$ for different $L$ cross together at $K=K_c$, which indicates the delocalization-MBDL transition. To determine the MBDL transition point $K_c$, we collapse these curves in Fig. \ref{figS2}(a2) by considering $S_F/L$ as a function of $(K-K_c)L^{1/\nu}$ with $K_c\approx0.25$. For other values of $J=\{1, 1.5, 2.0, 2.5\}$ in Figs. \ref{figS2}(b1-e1), the corresponding critical points can be determined as $K_c\approx\{0.40,0.51,0.62,0.74\}$ in Figs. \ref{figS2}(b2-e2), respectively. The result of MBDL transition point $\{K_c\}$ are in consistent with ones obtained from $O_F$, while the critical exponent $\nu\approx0.65$ in Figs. \ref{figS2}(a2-e2) is slightly different with that in Figs. \ref{figS3}(a2-e2).

Figures \ref{figS3}(a1-e1) show the result of $O_F$ versus the kick strength $K$ for various system sizes $L$ and interaction strengths $J$. For the case of $J=0.5$ in Fig. \ref{figS3}(a1), $O_F$ saturates to small finite value at weak kick strength, which indicates the thermalization of the systems. As the kick strength is increased, the curves of $O_F$ is separated around $K=K_c$ and increases with $L$. The curves can collapse well by considering $O_F$ as a function of $(K-K_c)L^{1/\nu}$ with the critical exponent $\nu\approx0.70$ and $K_c\approx0.2$, as shown in Fig. \ref{figS3}(a2), which indicates a delocalization-MBDL transition when approaching to the thermodynamic limit. Similar finite-size scaling analysis are taken for $J=\{1, 1.5, 2.0, 2.5\}$ in Figs. \ref{figS3}(b1-e1), which give the critical points at $K_c\approx\{0.38,0.52,0.62,0.75\}$ in Figs. \ref{figS3}(b2-e2).

\begin{figure*}[tb]
	\centering
	\includegraphics[width=0.6\textwidth]{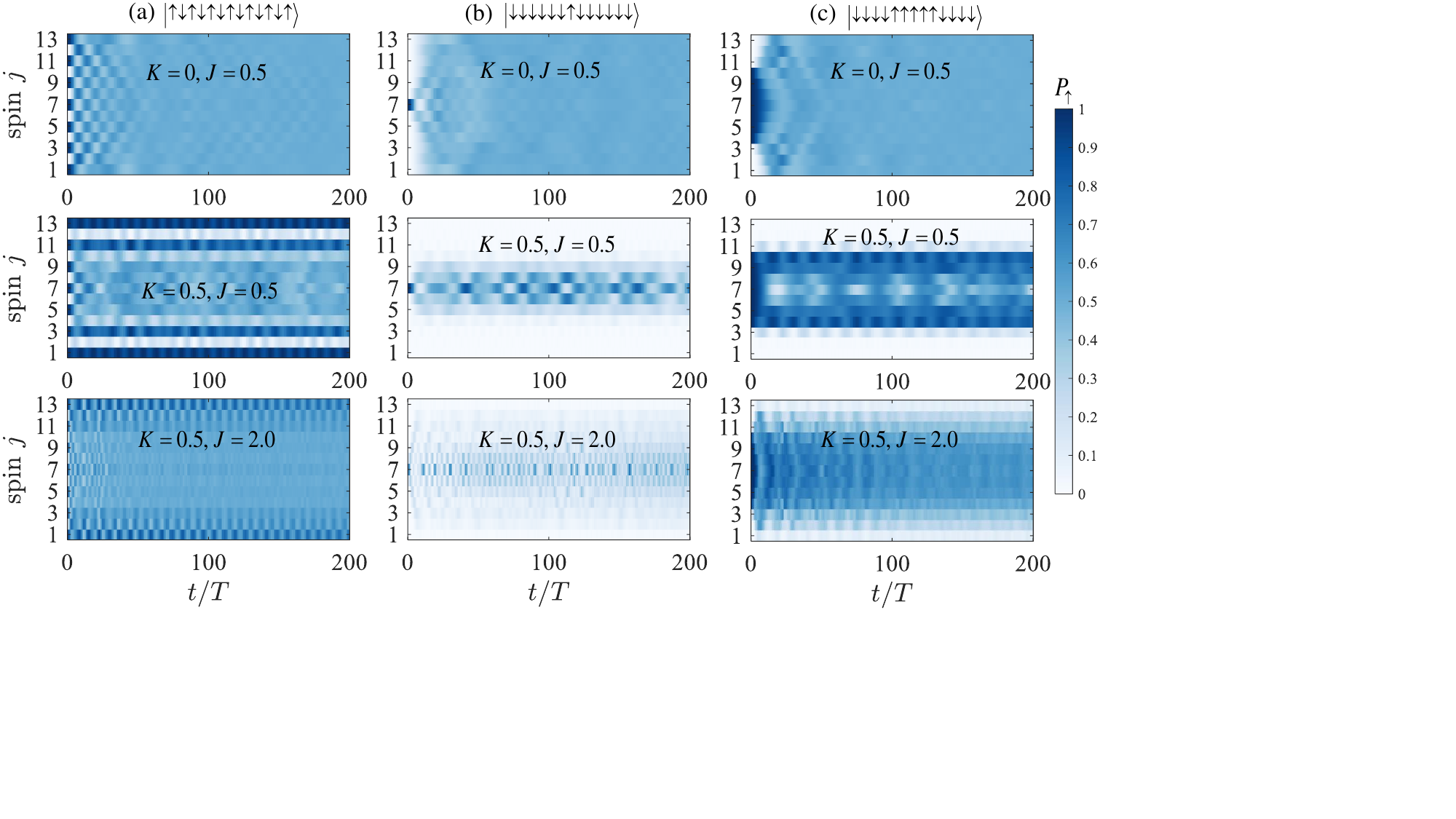}
	\caption{(Color online) Time evolution of the spin-up probability distribution $P_{\uparrow}(j,t)$ for (a) 
N\'{e}el-type state; (b) 
single-excitation state; and (c) %
domain-wall state. The parameters in the figures from upper to lower rows are $(K/\Omega,J/\Omega)=(0,0.5)$, $(0.5,0.5)$ and $(0.5,2.0)$, respectively. Independence of initial states, they exhibit the delocalization (thermalization) dynamics in static case without kicks, the kick-induced MBDL dynamics, and interaction-driven delocalization dynamics, respectively.
	}\label{figS4}
\end{figure*}

\subsection{C. Dynamics of various initial states}

The (de)localization of the many-body Floquet eigenstates $\{\ket{\psi^F_{\alpha}}\}$ can be observed through dynamics of spin-up (excitation) density distribution $P_{\uparrow}(j,t)$ defined in main text. The results of $P_{\uparrow}(j,t)$ as functions of $t/T$ for different initial states with different parameters $(K/\Omega,J/\Omega)$ are showed in Fig. \ref{figS4}. For N\'eel state $\ket{\psi(0)}=\ket{\uparrow\downarrow\uparrow\downarrow\uparrow\downarrow\uparrow\downarrow\uparrow\downarrow\uparrow\downarrow\uparrow}$ with $(K/\Omega,J/\Omega)=(0,0.5)$ in Fig. \ref{figS4}(a) , Hamiltonian Eq. 1 reduces to $H_{XY}$. In this case, the system tends to thermalize with losing embedded quantum information [the upper panel]. Turning on kick strength up to MBDL regime with $K=J=0.5$, the system turns to preserve the quantum coherence of initial states without thermalization in a long-time evolution up to $t=200T$ [the middle panel]. This disorder-free non-thermalized dynamical phase comes from dynamical constraints induced by quadratic kick potential. Noted that the system slightly thermalize around the central spin, which is due to the zero kick strength on spin $j=7$. Further increasing the interaction strength  up to $J=2.0$, the systems reenter the delocalized regime of phase diagram, which leads to delocalization dynamics of spin chains [the lower panel]. Similar phenomenons can be observed for other typical initial states in Fig. 1(b, c), which indicates the independence of the choice of initial states for observing MBDL and delocalization dynamics.

\begin{figure*}[tb]
	\centering
	\includegraphics[width=0.9\textwidth]{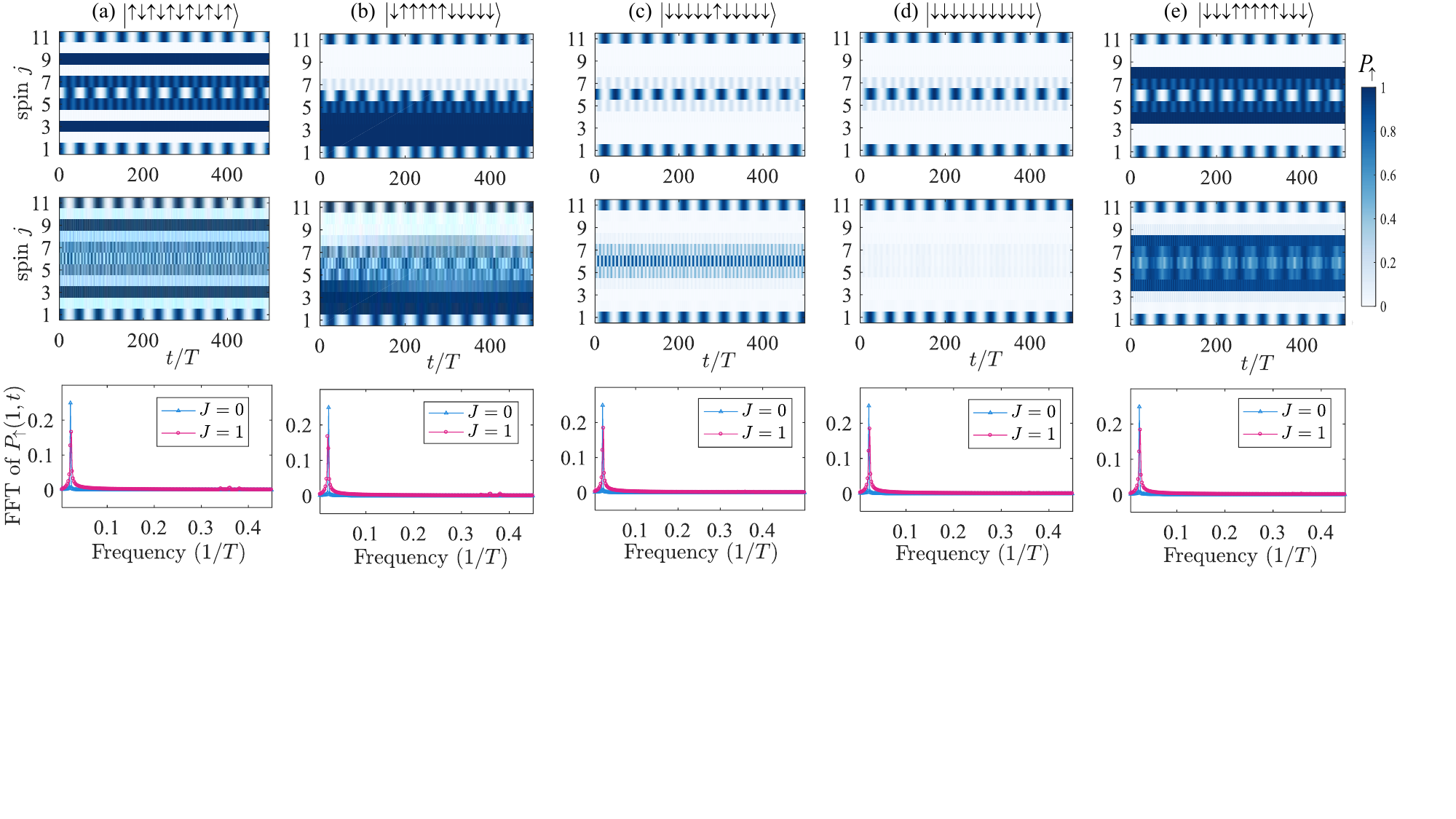}
	\caption{(Color online) Time evolution of the spin-up probability distribution $P_{\uparrow}(j,t)$ for (a)
N\'{e}el-type state; (b)
high-temperature state; (c) 
single-excitation state; (d) 
vacuum state; and (e) 
domain-wall state.
The figures in the upper and center rows show the spin dynamics in the non-interacting case of $J=0$ and in the interacting case of $J/\Omega=1$ with fixed $K/\Omega=16\pi/25$, respectively. The figures in the lowest row show the Fourier spectra of edge-spin oscillations for $J=0,1$ in the corresponding upper and center rows, which indicate the decoupling of edge spins with Rabi-like oscillations with frequency $\omega_d/\Omega=1/16\pi$ for different initial states.
	}\label{figS5}
\end{figure*}

The edge spins of the spin chains exhibit persistent Rabi oscillation in $P_{\uparrow}(j,t)$ for specific value of kick strength. Such phenomenon is induced by DD of edge spins in MBDL regime, which is independent of the choice of initial state. Here we consider systems size $L=11$ and kick strength $K=K_d^{(1,1)}$ to enable the DD of edge spins. The coherent dynamics and DD of edge spins of N\'eel state $\ket{\psi(0)}=\ket{\uparrow\downarrow\uparrow\downarrow\uparrow\downarrow\uparrow\downarrow\uparrow\downarrow\uparrow\downarrow\uparrow}$ for $J=0,1$ are shown in upper and middle panels of Fig. \ref{figS5}(a), respectively. One can see the Rabi oscillation of $P_{\uparrow}(1,t)$ and $P_{\uparrow}(11,t)$ independent of interaction strength $J=0,1$. By applying the Fourier transformation of $P_{\uparrow}(1,t)$ for varying $J=0,1$, we obtain the corresponding Fourier spectrum in Fig. \ref{figS5}(a) [lower panel]. The Rabi-like oscillation frequency of $P_{\uparrow}(1,t)$ for $J=1$ are the same as that for free spins with $J=0$. This indicates that the spins of $j=1,L$ are effectively decoupled and behavior as free spins with Rabi oscillation. Noted that Rabi oscillation also exhibits in central spin $j=6$ for decoupled limit $J=0$, which breaks up when interaction strength dominates the dynamics around central spins with $J=1$. DD of other initial state can also be observed in Fig. \ref{figS5}(b-e).

\begin{figure}[tb]
\centering
\includegraphics[width=0.45\textwidth]{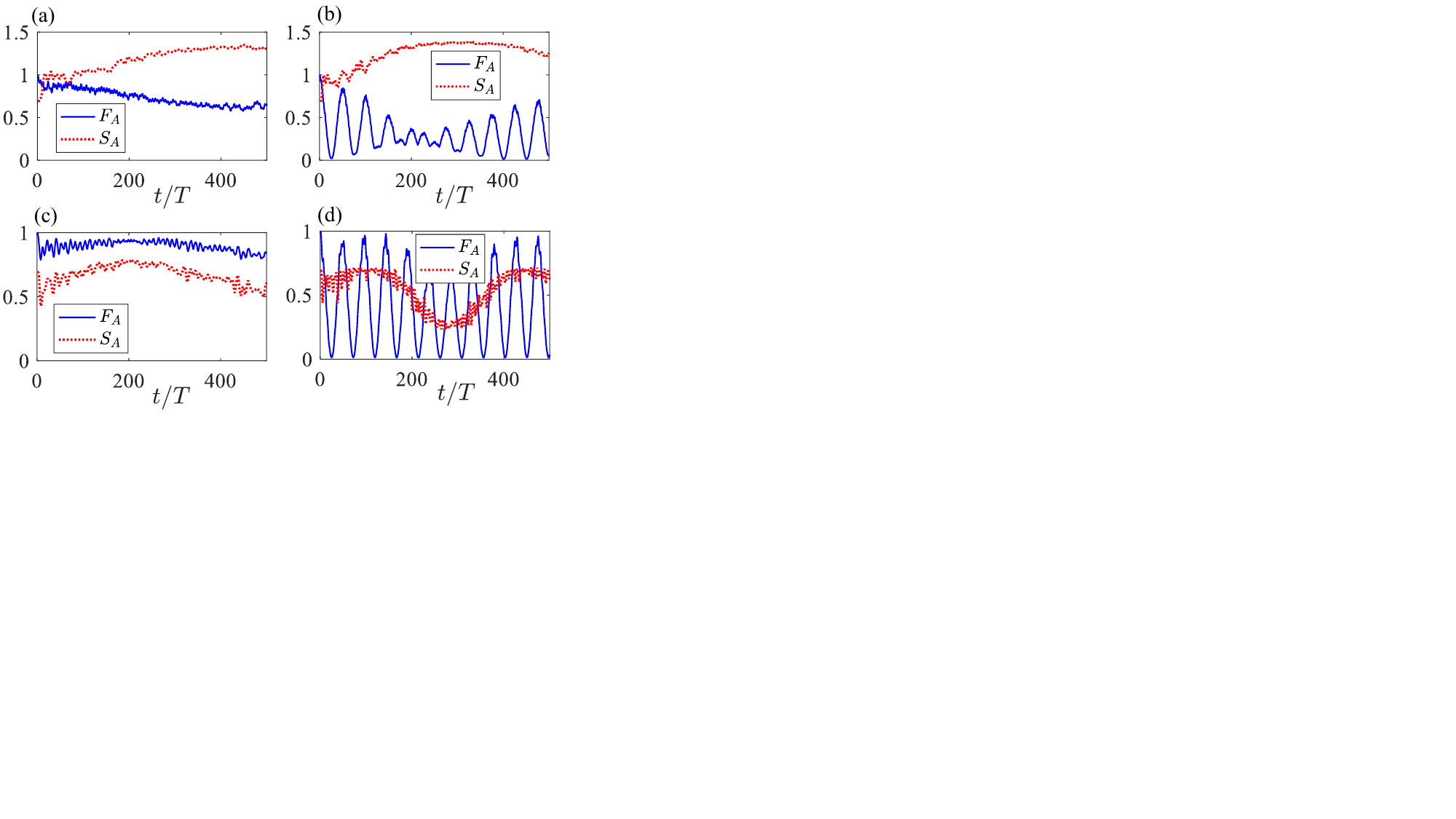}
\caption{(Color online) Dynamics of the subsystem fidelity $F_A$ and entanglement entropy $S_A$ from (a,b) high-temperature Bell state $\ket{\psi(0)}=\frac{1}{\sqrt{2}}(\ket{\downarrow\uparrow\uparrow\uparrow\uparrow\uparrow\downarrow\downarrow\downarrow\downarrow\downarrow} +\ket{\uparrow\downarrow\downarrow\downarrow\downarrow\downarrow\uparrow\uparrow\uparrow\uparrow\uparrow})$; and from (c,d) entangled state $\ket{\psi(0)}=\frac{1}{\sqrt{2}}(\ket{\uparrow}_4\ket{\downarrow}_5+\ket{\downarrow}_4\ket{\uparrow}_5)\otimes\prod_{j\neq4,5}\ket{\downarrow}_j$.
The system size is $L=11$ and the parameters are $(K/\Omega,J/\Omega)=(1.0,0.5)$ in (a,c) and $(16\pi/25,1.0)$ in (b,d), respectively.
}\label{figS6}
\end{figure}

The persistent coherent dynamics protected by the disorder-free MBDL is exhibited for generic initial states. We compute the time evolution of fidelity $F_A(t)$ and entanglement entropy $S_A(t)$ of the high-temperature Bell states and entangled states in Figs. \ref{figS6}(a,b) and (c,d), respectively. The parameters are $(K/\Omega,J/\Omega)=(1.0,0.5)$ and $(16\pi/25,1.0)$ in (a,c) and (b,d), respectively. The subsystem fidelity dynamics with low entanglement entropy persists in the long-time evolution up to $t=500T$.
Noted that $F_A(t)$ oscillates in (b,d) due to DD of edge spins.

\begin{figure}[tb]
	\centering
	\includegraphics[width=0.45\textwidth]{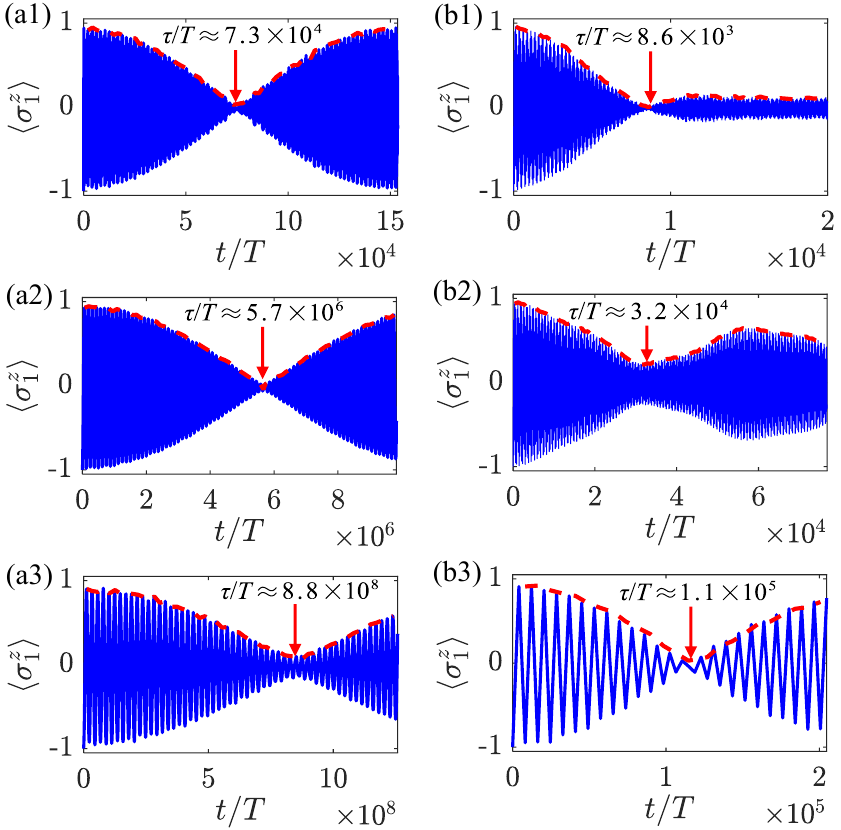}
	\caption{(Color online) Dynamics of $\braket{\sigma_1^z}$ starting from (a1,a2,a3) vacuum states $\ket{\psi(0)}=\ket{\downarrow\downarrow\cdot\cdot\cdot}$; and from (b1,b2,b3) entangled states $\ket{\psi(0)}=\frac{1}{\sqrt{2}}(\ket{\uparrow}_4\ket{\downarrow}_5+\ket{\downarrow}_4\ket{\uparrow}_5)\otimes\prod_{j\neq4,5}\ket{\downarrow}_j$ under the local DD effect. Red dashed lines denote upper envelop of $\braket{\sigma_1^z}$. Lifetime $\tau/T$ are roughly determined by the first minimum point of the envelop, with the values being plotted in the figure. The system sizes are (a1,b1) $L=7$, (a2,b2) $L=9$, and (a3,b3) $L=11$, and the parameters are $(K/\Omega,J/\Omega)=(K_d^{(1,1)},1)$ for ensuring the DD of edge spins.
	}\label{figS7}
\end{figure}

We can also estimate the lifetime $\tau$ of the persist Rabi-like oscillation of edge spins from the long-time evolution of $\braket{\sigma_1^z}$ under the DD effect. For instance, we show the time evolution of $\braket{\sigma_1^z}$ in Figs. \ref{figS7}(a,b,c) starting from the vacuum states $\ket{\psi(0)}=\ket{\downarrow\downarrow\cdot\cdot\cdot}$ and (d,e,f) for entangled states $\ket{\psi(0)}=\frac{1}{\sqrt{2}}(\ket{\uparrow}_4\ket{\downarrow}_5+\ket{\downarrow}_4\ket{\uparrow}_5)\otimes\prod_{j\neq4,5}\ket{\downarrow}_j$, for system sizes $L=7,8,11$. The parameters $(K/\Omega,J/\Omega)=(K_d^{(1,1)},1)$ are set to enable the DD of edge spins for various $L$. The upper envelop of $\braket{\sigma_1^z}$ denoted by red dashed lines decrease to local minimum with increasing time. The lifetime are roughly determined by the first minimum point of the envelop. The obtained values of lifetime $\tau/T$ are plotted in the figure.

\begin{figure}[tb]
	\centering
	\includegraphics[width=0.45\textwidth]{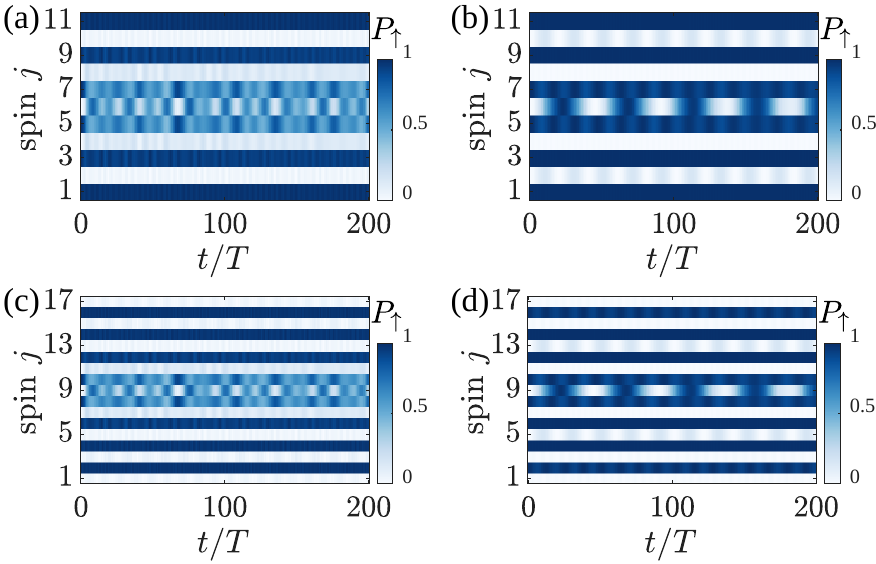}
	\caption{(Color online) Time evolution of the spin-up probability distribution $P_{\uparrow}(j,t)$ for (a,b) $L=11$ and (c,d) $L=17$. Other parameters fixed for (a,c) $(J/\Omega,K/\Omega)=(0.5,1.5)$ and (b,d) $(J/\Omega,K/\Omega)=(0.1,3)$.
	}\label{figS8}
\end{figure}

Noted that some spins in the cental region of the chain could be non-thermalized and then partially loss the initial-state information even in the MBDL region, due to the quadratic form of the periodic kick potential, such as those in Fig. \ref{figS4}. To improve the fidelity of initial spin states for quantum storage, one can increase the ratio between the kick and interaction strengths $K/J$. For instance, the center three spins in the chain of $L=11$ becomes non-thermalized when the ratio $K/J=3$ in Fig. \ref{figS8}(a) is increased to $K/J=30$ in Fig. \ref{figS8}(b). Similar results for a large spin chain of $L=17$, where more spins are dynamically decoupled, are shown in Figs. \ref{figS8}(c,d).

\subsection{D. Derivation of the relation between $K$ and $N_d$}
There may exist more than two DD spins fulfill the condition $K_d^{(j,m)}$ at the same time. For example,
\begin{equation}\label{K1}
	\begin{aligned}
	    &K_d^{(j_0+1,1)}=K_d^{(j_0+2,4)}=
		\cdots=K_d^{(j_0+j_1,j_1^2)}=\frac{\pi}{T},\\
		&K_d^{(j_0+2,1)}=K_d^{(j_0+4,4)}=
		\cdots=K_d^{(j_0+j_2,j_2^2/4)}=\frac{\pi}{4T},
	\end{aligned}
\end{equation}
where $j_1=(L-1)/2$ and $j_2=2\lfloor(L-1)/4\rfloor$. Noted that the total number of DD spins $N_d$ depends on the number of the total terms in Eq. (\ref{K1}), one can obtain
\begin{equation}
	\begin{aligned}
		&N_d^{(1)}=L-1,\\
		&N_d^{(2)}=2\lfloor(L-1)/4\rfloor,
	\end{aligned}
\end{equation}
respectively. The general formalization of Eq. (\ref{K1}) is
\begin{equation}
	 K_d^{(j_0+l,1)}=K_d^{(j_0+2l,4)}=\cdots=K_d^{(j_0+j_l,j_l^2/l^2)}=\frac{\pi}{l^2T},
\end{equation}
where $l\leq (L-1)/2$, $j_l=l\lfloor(L-1)/(2l)\rfloor$, hence
\begin{equation}
   	N_d^{(l)}=2j_l/l=2\lfloor(L-1)/(2l)\rfloor.
\end{equation}
By iterating $l$ from 1 to $(L-1)/2$, one can obtain all the values of $\{N_d^{(l)}\}$ with corresponding kick strength $K={\pi}/{l^2T}$, which leads to a relation between $N_d$ and $K$
\begin{equation}
	K=\frac{\pi}{T \lfloor(L-1)/N_d\rfloor ^2}.
\end{equation}

\end{document}